# Performance of Optimal Data Shaping Codes

**Yi Liu**\*, **Pengfei Huang**\*, and **Paul H. Siegel**\*

\*Electrical and Computer Engineering Dept., University of California, San Diego, La Jolla, CA 92093 U.S.A
{*yil333, pehuang, psiegel*}@ucsd.edu

*Abstract*—Data shaping is a coding technique that has been proposed to increase the lifetime of flash memory devices. Several data shaping codes have been described in recent work, including endurance codes [2] and direct shaping codes for structured data [8], [4], [5]. In this paper, we study information-theoretic properties of a general class of data shaping codes and prove a separation theorem stating that optimal data shaping can be achieved by the concatenation of optimal lossless compression with optimal endurance coding. We also determine the expansion factor that minimizes the total wear cost. Finally, we analyze the performance of direct shaping codes and establish a condition for their optimality.

## I. INTRODUCTION

NAND flash memory has become a ubiquitous data storage technology. It uses rectangular arrays, or *blocks* of floating-gate transistors (commonly referred to as *cells*) to store information. The flash memory cells gradually wear out with repeated writing and erasing, referred to as *program/erase (P/E) cycling*. The damage caused by P/E cycling is dependent on the programmed cell level. For example, in SLC flash memory, each cell has two possible levels, erased and programmed, represented by 1 and 0, respectively. Storing 1 in a cell causes less damage, or *wear*, than storing 0. More generally, in multilevel flash memories, the cell wear is an increasing function of the threshold voltage associated with the programmed cell levels. Several coding techniques have been proposed to increase the lifetime of flash memory device by reducing the total cost wear to the flash device.

*Endurance coding*, intended for statistical shaping of random data, was proposed in [2]. The objective is to reduce wear by converting unconstrained data to encoded sequences with a prescribed distribution of cell levels. For a given cell-level "wear-cost" model and a specified target code rate, the optimal distribution of levels that minimizes the average wear-cost was determined analytically. Theoretically achievable gains for SLC and 4-level MLC devices were also computed. *Adaptive endurance coding* (AEC) for structured source data, which uses a concatenation of lossless data compression with endurance coding, was also proposed. The performance of enumerative endurance codes that reduce the average fraction of 0 symbols was evaluated empirically on SLC devices, as well as on MLC devices (applied to both pages). Results of a device-level simulation of AEC on SLC flash were also presented. A realization of endurance coding was presented in [3].

In [8], low-complexity, rate-1, *direct shaping codes* for structured data were proposed for use on SLC flash memory. During encoding, a rate-1 mapping dictionary is adaptively generated to efficiently transform a structured data sequence directly into a sequence that induces less cell wear. This code construction was extended to an MLC direct shaping code in [4], [5]. Computer simulation was used to evaluate the performance of a direct shaping code on English language text. Error propagation properties of direct shaping codes were also analyzed.

In this paper, we study the performance of a general class of data shaping codes. In Section II, we use known properties of word-valued sources to determine the symbol occurrence probability of shaping codeword sequences. In Section III, we develop a theoretical bound on the trade-off between the rate and the average wear cost of a shaping code. A consequence of the analysis is a separation theorem, which states that optimal shaping can be achieved by a concatenation of lossless compression and endurance coding. The expansion factor that minimize total wear cost is also determined. We then specialize the analysis to SLC direct shaping codes. We show that they are in general sub-optimal as rate-1 shaping codes, and provide necessary and sufficient source conditions for their optimality.

## II. INFORMATION-THEORETIC PRELIMINARIES

### A. Basic Model

First, we fix some notation. Let $\mathbf{X} = X_1 X_2 \ldots$, where $X_i \sim X$ for all $i$, be an i.i.d source with alphabet $\mathcal{X} = \{\alpha_1, \ldots, \alpha_u\}$. We use $|\mathcal{X}|$ to denote the size of the alphabet, $|x^*|$ to denote the length of sequence $x^*$ and use $P(x^*)$ to denote the probability of any finite sequence $x^*$. Let $\mathcal{Y} = \{\beta_1, \ldots, \beta_v\}$ be an alphabet and $\mathcal{Y}^*$ be the set of all finite sequences over $\mathcal{Y}$, including the null string $\lambda$ of length 0. Each $\beta_i$ is associated with a wear cost $U_i$. Without loss of generality, we assume that $U_1 \leqslant U_2 \leqslant \ldots U_v$. We use a cost vector $\mathcal{U} = [U_1, U_2, \ldots, U_v]$ to represent the wear cost associated with alphabet $\mathcal{Y}$.

A general shaping code is defined as a prefix-free mapping $\phi : \mathcal{X}^q \to \mathcal{Y}^*$ which maps a length-$q$ data sequence $x_1^q$ to a variable length code sequence $y^*$. We use $\mathbf{Y}$ to denote the process $\phi(\mathbf{X}^q)$, where $\mathbf{X}^q$ is the vector process $X_1^q, X_{q+1}^{2q}, \ldots$. The entropy rate of the process $\mathbf{Y}$ is

$$H(\mathbf{Y}) = \lim_{n \to \infty} \frac{1}{n} H(Y_1 Y_2 \ldots Y_n). \tag{1}$$

We denote the length of a codeword $\phi(x_1^q)$ by $L(\phi(x_1^q))$ and the expected length of codewords generated by length-$q$ source sequences is given by

$$E(L) = \sum_{x_1^q \in \mathcal{X}^q} P(x_1^q) L(\phi(x_1^q)). \tag{2}$$

The *expansion factor* is defined as the ratio of the codeword expected length to the length of the input source sequence, namely

$$f = E[L]/q. \tag{3}$$

**Remark 1.** Endurance codes and direct shaping codes can be treated as special cases of this general class of shaping codes. Endurance codes are used when the source has a uniform distribution, with entropy rate $H(\mathbf{X}) = \log_2 |\mathcal{X}|$. A length-$m$ direct shaping code is a shaping code with $q = 1$, $f = 1$, where both $\mathbf{X}$ and $\mathbf{Y}$ have alphabet size $2^m$.

The pair $\mathbf{X}$ and $\phi$ form a word valued source, as defined in [7]. The following theorem, proved in [7], gives the entropy rate of the shaping code $\phi(\mathbf{X}^q)$.

**Theorem 1.** *For a prefix-free shaping code $\mathbf{Y} = \phi(\mathbf{X}^q)$ such that $H(\mathbf{X}^q) < \infty$ and $E(L) < \infty$, the entropy rate of the encoder output satisfies*

$$H(\mathbf{Y}) = \frac{H(\mathbf{X}^q)}{E(L)} = \frac{qH(\mathbf{X})}{E(L)}. \quad (4)$$

$\square$

### B. Asymptotic Symbol Occurrence Probability

For simplicity, and without loss of generality, we assume $q = 1$. The mapping is $\phi : \mathcal{X} \to \mathcal{Y}^*$. Let $y_1^l$ denote the first $l$ symbols of $\phi(\mathbf{X})$. We assume the wear cost is independent and additive, so the wear cost of sequence $y_1^l$ can be expressed as

$$W(y_i^l) = \sum_{i=1}^{v} N_i(y_1^l) U_i \quad (5)$$

where $N_i(y_1^l)$ stands for the number of occurrences of $\beta_i$ in sequence $y_1^l$. The wear cost per code symbol is therefore $\sum_i N_i(y_1^l) U_i / l$.

Let

$$Q(y_1^l) = Pr\{Y_1^l = y_1^l\} \quad (6)$$

denote the probability distribution of $Y_1^l$. The expected wear cost per symbol of a length-$l$ shaping code sequence is

$$W_l = \sum_{y_1^l \in \mathcal{Y}^l} Q(y_1^l) W(y_1^l) / l$$
$$= \sum_{i=1}^{v} \sum_{y_1^l \in \mathcal{Y}^l} Q(y_1^l) N_i(y_1^l) U_i / l. \quad (7)$$

The asymptotic expected wear cost per symbol, or *average wear cost* of a data shaping code is

$$A(\phi(\mathbf{X})) = \lim_{l \to \infty} W_l. \quad (8)$$

Let

$$P_i = \lim_{l \to \infty} \sum_{y_1^l} Q(y_1^l) N_i(y_1^l) / l = \lim_{l \to \infty} \frac{E(N_i(Y_1^l))}{l} \quad (9)$$

be the asymptotic probability of occurrence of symbol $\beta_i$. Clearly, $\sum_i P_i = 1$. Then the average cost of a data shaping code can be expressed as

$$A(\phi(\mathbf{X})) = \sum_i P_i U_i. \quad (10)$$

In the rest of this subsection, we will show how to calculate $P_i$. Define the prefix operator $\pi$ as $y_1^n \pi^i = y_1^{n-i}$ for $0 \le i < n$ and $y_1^n \pi^i = \lambda$ for $i \ge n$. Let $\pi\{y^*\}$ denote the set of all the prefixes of a sequence $y^*$. We denote by $\mathcal{G}_\phi(y_1^l)$ the set of all sequences $x^* \in \mathcal{X}^*$ such that $y_1^l$ is a prefix of $\phi(x^*)$ but not of $\phi(x^*\pi)$. That is,

$$\mathcal{G}_\phi(y_1^l) = \{x^* \in \mathcal{X}^* | y_1^l \in \pi\{\phi(x^*)\} \wedge |\phi(x^*\pi)| < l\} \quad (11)$$

and the distribution of $y_1^l$ can be expressed as

$$Q(y_1^l) = \sum_{x^* \in \mathcal{G}_\phi(y_1^l)} P(x^*). \quad (12)$$

We define by $M_l$ the minimum length of sequence $x_1^n$ such that $|\phi(x_1^n)| \ge l$ and let $S_{M_l}$ be the length of $\phi(x_1^{M_l})$. Note that

$$S_{M_l - 1} < l \le S_{M_l}. \quad (13)$$

According to [7], the random variable $M_l$ satisfies the property of being a *stopping rule* for the sequence of i.i.d. random variable $\{\phi(X^\infty)\}$. Wald's equality [9] then implies that

$$E(N_i(\phi(X_1^{M_l}))) = E(N_i(\phi(X))) E(M_l). \quad (14)$$

**Remark 2.** Given a nonnegative-valued function $f$, let $F_i = f(X_i)$ and assume $E(F) < \infty$. We should point out that $E(F_{M_l})$ is not always equal to $E(F)$. For details, see [7]. The following two lemmas were proved in [7].

**Lemma 2.** *Given a nonnegative-valued function $f$, let $F_i = f(X_i)$. If $E(F) < \infty$, then*

$$\lim_{l \to \infty} \frac{E(F_{M_l})}{l} = 0. \quad (15)$$

$\square$

**Lemma 3.** *If $E[L] < \infty$, then*

$$\lim_{l \to \infty} \frac{E[M_l]}{l} = \frac{1}{E(L)}. \quad (16)$$

$\square$

Using these results, we derive a lemma which tells us how to calculate the asymptotic occurrence probability of encoder output process $\mathbf{Y}$.

**Lemma 4.** *For a shaping code $\phi : \mathcal{X}^q \to \mathcal{Y}^*$, the occurrence probability $P_i$ of codeword sequence is given by*

$$P_i = Pr(\hat{Y} = \beta_i) = E(N_i(\phi(X^q))) \frac{1}{E(L)}. \quad (17)$$

*Proof:* See Appendix A. ∎

### C. Lower Bound on Symbol Distribution Entropy

The probability distribution of symbols determines the average wear cost of the shaping code. To find the minimum possible average wear cost, we first study some properties of the distribution $\{P_i\}$. For convenience, define the random variable $\hat{Y}$ by

$$Pr(\hat{Y} = \beta_i) = P_i. \quad (18)$$

The following lemma gives a lower bound on its entropy, $H(\hat{Y})$.

**Lemma 5.** *The entropy $H(\hat{Y})$ is lower bounded by the entropy rate of the shaping code sequence, i.e.,*

$$H(\hat{Y}) \ge H(\mathbf{Y}). \quad (19)$$

$\square$

*Proof:* The entropy rate of $\mathbf{Y}$ is

$$H(\mathbf{Y}) = \lim_{l \to \infty} \frac{1}{l} H(Y_1^l) \quad (20)$$

and we have

$$H(Y_1^l) \le \sum_{i=1}^{l} H(Y_i). \quad (21)$$

Random variable $Y_i$ has probability distribution

$$\Pr(Y_i = \beta_j) = \sum_{y_1^l : y_i = \beta_j} Q(y_1^l). \quad (22)$$

We use $P_{ij}$ to denote $\Pr(Y_i = \beta_j)$. Then $H(Y_i)$ can be expressed as

$$H(Y_i) = -\sum_{j=1}^{v} P_{ij} \log_2 P_{ij}. \quad (23)$$

Since the entropy function is a concave function, we have, by Jensen's inequality,

$$\sum_{i=1}^{l} H(Y_i) = -\sum_{i=1}^{l}\sum_{j=1}^{v} P_{ij} \log_2 P_{ij}$$
$$\leqslant -l \sum_{j=1}^{v} \left(\frac{\sum_i P_{ij}}{l}\right) \log_2\left(\frac{\sum_i P_{ij}}{l}\right) = l H(Y_l'), \quad (24)$$

where the random variable $Y_l'$ has probability distribution

$$\Pr(Y' = \beta_j) = \frac{1}{l}\sum_{i=1}^{l} P_{ij}. \quad (25)$$

Notice that

$$\frac{1}{l}\sum_{i=1}^{l} P_{ij} = \frac{1}{l}\sum_{i=1}^{l}\sum_{y_1^l : y_i = \beta_j} Q(y_1^l) = \frac{1}{l}\sum_{y_1^l}\sum_{i : y_i = \beta_j} Q(y_1^l). \quad (26)$$

For a specific length-$l$ sequence $y_1^l$, the quantity $Q(y_1^l)$ appears $N_j(y_1^l)$ times in the summation in (26). Therefore,

$$\frac{1}{l}\sum_{i=1}^{l} P_{ij} = \frac{1}{l}\sum_{y_1^l} N_j(y_1^l) Q(y_1^l)$$
$$= \frac{E(N_j(Y_1^l))}{l}. \quad (27)$$

This implies that $Y_l'$ converges in distribution to $\hat{Y}$ and

$$\lim_{l \to \infty} H(Y') = H(\hat{Y}). \quad (28)$$

Combining equations (20), (21), (24), and (28), we have

$$H(\mathbf{Y}) = \lim_{l \to \infty} \frac{1}{l} H(Y_1^l)$$
$$\leqslant \liminf_{l \to \infty} \frac{1}{l}\sum_{i=1}^{l} H(Y_i) \leqslant \lim_{l \to \infty} H(Y') = H(\hat{Y}). \quad (29)$$

∎

**Remark 3.** From the proof, we can conclude that if $H(\hat{Y})$ equals $H(\mathbf{Y})$, then the following two conditions must hold.
(a) $H(Y_1^l) = \sum_{i=1}^{l} H(Y_i)$ for any $l$; i.e., the random variables $Y_i$ are mutually independent.
(b) $\sum_{i=1}^{l} H(Y_i) = l H(Y_l')$ for any $l$; i.e., the random variables $Y_i$ are identically distributed.

These two conditions, in turn, imply that $Y_1 Y_2 \ldots$ acts like an i.i.d sequence generated by $\hat{Y}$.

**Example 1.** Consider the uniformly distributed source $\mathbf{X}$ and the shaping code defined by the mapping $\{00 \to 000, 01 \to 001, 10 \to 01, 11 \to 1\}$. The occurrence probability of symbol 0 is $2/3$ and

$$H(\hat{Y}) = -\frac{1}{3}\log_2\frac{1}{3} - \frac{2}{3}\log_2\frac{2}{3} \simeq 0.9183. \quad (30)$$

The entropy rate of the shaping code sequence is

$$H(\mathbf{Y}) = \frac{H(\mathbf{X}^2)}{E(L)} = \frac{2}{2.25} = 0.8889. \quad (31)$$

We see that $H(\mathbf{Y}) < H(\hat{Y})$.

### III. OPTIMAL SHAPING CODES

*A. Cost Minimizing Probability Distribution*

Given a data shaping code $\phi : \mathcal{X}^q \to \mathcal{Y}^*$, assume that after $nq$ source symbols are encoded, the codeword sequence is $\phi(x_1^{nq})$. As in (7), (8), (9), we define the expected wear cost per *source* symbol, or *total wear cost* of a data shaping code as

$$T(\phi(\mathbf{X}^q)) = \frac{\sum_i E(N_i(\phi(X^{nq}))) U_i}{nq} = \frac{\sum_i E(N_i(\phi(X^q))) U_i}{q}$$
$$= \frac{E(L)}{q}\frac{\sum_i E(N_i(\phi(X^q))) U_i}{E(L)} = f\sum_{i=1}^{v} P_i U_i. \quad (32)$$

To increase the lifetime of a flash memory device, we want to find the data shaping code that can minimize the total wear cost. This is equivalent to solving the following optimization problem.

$$\begin{aligned}
\underset{P_i', f}{\text{minimize}} \quad & f\sum_{i=1}^{v} P_i' U_i \\
\text{subject to} \quad & H(\hat{Y}) \geqslant H(\mathbf{Y}) = \frac{H(\mathbf{X})}{f} \quad (33) \\
& \sum_i P_i' = 1.
\end{aligned}$$

In order to do that, we first prove a lower bound on the minimum average wear cost achievable with a fixed expansion factor $f$.

**Theorem 6.** *Given the distribution $P$ of source words and a cost vector $\mathcal{U}$, the average cost of a shaping code $\phi : \mathcal{X}^q \to \mathcal{Y}^*$ with expansion factor $f$ is lower bounded by $\sum_i P_i' U_i$, with $P_i' = \frac{1}{N} 2^{-\mu U_i}$, where $N$ is a normalization constant such that $\sum_i P_i' = 1$ and $\mu$ is a positive constant such that $\sum_i -P_i' \log P_i' = H(\mathbf{X})/f$.* □

*Proof:* From Theorem 1 and Lemma 5, we see that, for a shaping code $\phi$ with expansion factor $f$, the following inequality holds:

$$H(\hat{Y}) \geqslant H(\mathbf{Y}) = \frac{qH(\mathbf{X})}{E(L)} = \frac{H(\mathbf{X})}{f}. \quad (34)$$

To calculate the minimum possible average cost, we must solve the optimization problem.

$$\begin{aligned}
\underset{P_i'}{\text{minimize}} \quad & \sum_i P_i' U_i \\
\text{subject to} \quad & H(\hat{Y}) \geqslant \frac{H(\mathbf{X})}{f} \quad (35) \\
& \sum_i P_i' = 1.
\end{aligned}$$

We divide this optimization problem into two parts. First, we fix $H(\hat{Y})$ and find the optimal symbol occurrence probabilities. Then we find the optimal $H(\hat{Y})$ to minimize the average wear cost. The optimization problem then becomes

$$\begin{aligned}\underset{H(\hat{Y})}{\text{minimize}}\quad &\underset{P'_i}{\text{minimize}}\quad \sum_i P'_i U_i\\ \text{subject to}\quad &H(\hat{Y}) \geqslant \frac{H(\mathbf{X})}{f} \qquad (36)\\ &\sum_i P'_i = 1.\end{aligned}$$

If we fix $H(\hat{Y})$, we can solve the optimization problem by using the method of Lagrange multipliers. The solution is

$$P'_i = \frac{1}{N} 2^{-\mu U_i} \qquad (37)$$

where $N = \sum_i 2^{-\mu U_i}$ is a normalization constant and $\mu$ is a non-negative constant such that

$$H(\hat{Y}) = \sum_i -P'_i \log_2 P'_i. \qquad (38)$$

Note that $\mu = 0$ if and only if $H(\hat{Y}) = \log_2 |\mathcal{Y}|$. For simplicity, let $h$ denote $H(\hat{Y})$. Then $\mu$ and $N$ are functions of $h$, which we denote by $N \stackrel{\text{def}}{=} N(h)$ and $\mu \stackrel{\text{def}}{=} \mu(h)$, respectively. Let $C(h) = \sum_i \frac{U_i}{N(h)} 2^{-\mu(h)U_i}$ be the minimum cost, given that $h \geqslant \frac{H(\mathbf{X})}{f}$. From (38), we see that

$$C(h) = \frac{h - \log_2 N}{\mu}. \qquad (39)$$

Then

$$\frac{dC}{dh} = \frac{\mu(1 - \frac{1}{N\ln 2}\frac{dN}{dh}) - (h - \log_2 N)\frac{d\mu}{dh}}{\mu^2}, \qquad (40)$$

where, as is easily verified,

$$\frac{dN}{dh} = -\sum_i U_i 2^{-\mu U_i} \frac{d\mu}{dh} \ln 2 = -N \ln 2 \frac{d\mu}{dh} C. \qquad (41)$$

Combining (40) and (41), we get

$$\frac{dC}{dh} = \frac{1}{\mu} + \frac{\mu C + \log_2 N - h}{\mu^2} \frac{d\mu}{dh}. \qquad (42)$$

Since $\mu C + \log_2 N = h = H(\hat{Y})$, we deduce that $\frac{dC}{dh} = \frac{1}{\mu} > 0$. Therefore, the minimum cost is achieved when $h = H(\hat{Y}) = \frac{H(\mathbf{X})}{f}$. ■

**Remark 4.** If the source $\mathbf{X}$ has a uniform distribution, then $\mu$ satisfies $-f \sum_i P'_i \log P'_i = \log_2 |\mathcal{X}|$. Thus, we recover the result in [2] characterizing endurance codes with minimum average wear cost.

**Remark 5.** When the minimum average wear cost is achieved, we have $H(\hat{Y}) = H(\mathbf{Y})$. Thus, the codeword sequence looks like an i.i.d sequence generated by $\hat{Y}$ (see Remark 3).

Using Theorem 6, we can calculate the total wear cost as a function of the expansion factor $f$. Fig. 1 shows the total wear cost curve for a quaternary source and code alphabet, a uniformly distributed source $\mathbf{X}$, and cost vector $\mathcal{U} = [1,2,3,4]$. There is an optimal value of $f$ and corresponding minimum total wear cost.

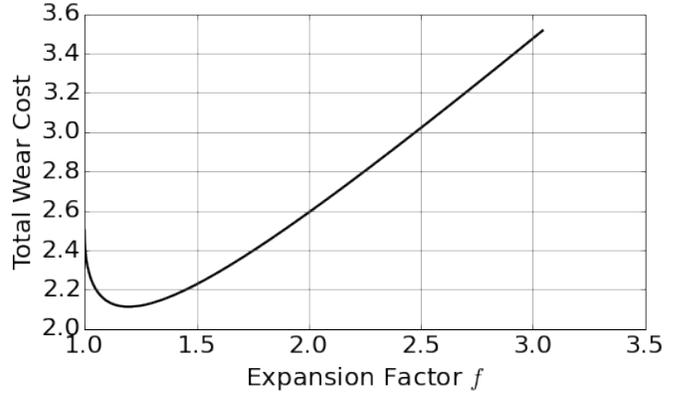

Fig. 1. Total wear cost versus $f$ for random source with $\mathcal{U} = [1,2,3,4]$

### B. Separation Theorem

We now prove a separation theorem for data shaping codes. This result shows that the lower bound on the average wear cost can be achieved by combining optimal lossless compression with optimal endurance coding. From this result and Theorem 6, we will then determine the expansion factor $f$ that achieves the minimum total wear cost.

**Theorem 7.** *For a given expansion factor $f$, the minimum average wear cost can be achieved by a concatenation of an optimal lossless compression code with an optimal endurance code.* □

*Proof:* Fix the expansion factor $f$. Consider a source $\mathbf{X}$ with alphabet $\mathcal{X}$. According to Shannon, the optimal asymptotic compression rate $g$ is given by

$$g = \frac{H(\mathbf{X})}{\log_2 |\mathcal{X}|}. \qquad (43)$$

From [7] we know that the probability distribution of the compressed sequences tends to a uniform distribution. If we now apply endurance coding to the compressed sequence, the expansion factor of the endurance code is

$$f' = \frac{f}{g} = \frac{f \log_2 |\mathcal{X}|}{H(\mathbf{X})}. \qquad (44)$$

Minimization of the average cost translates to solving the following optimization problem:

$$\begin{aligned}\underset{P'_i}{\text{minimize}}\quad &\sum_i P'_i U_i\\ \text{subject to}\quad &H(\hat{Y}) = \frac{\log_2 |\mathcal{X}|}{f'} = \frac{H(\mathbf{X})}{f} \qquad (45)\\ &\sum_i P'_i = 1.\end{aligned}$$

Comparing (45) to (36), we conclude that the minimum cost can be achieved by a shaping code that first applies optimal lossless compression to the source data, followed by optimal endurance coding. ■

We now use the separation theorem to determine the minimal achievable total wear cost of a data shaping code.

**Theorem 8.** *Let $P$ be the source distribution and let $\mathcal{U}$ be a cost vector. If $U_1 \neq 0$, then the minimum total wear cost of a shaping code $\phi : \mathcal{X}^q \to \mathcal{Y}^*$ is given by $f \sum_i P'_i U_i$, where $P'_i = 2^{-\mu U_i}$, $\mu$ is a positive constant selected such that $\sum_i 2^{-\mu U_i} = 1$, and the expansion factor $f$ is*

$$f = \frac{H(\mathbf{X})}{-\sum_i P'_i \log_2 P'_i}. \tag{46}$$

*If $U_1 = 0$, then the total cost is a decreasing function of $f$.* □

*Proof:* The proof is summarized below. See Appendix B for detailed calculations. The separation theorem implies that, for a source $\mathbf{X}$, we can first apply an optimal lossless compression to get a uniform source $\mathbf{X}'$. Now we consider the optimal $f'$ when endurance coding is applied to $\mathbf{X}'$. The total wear cost is $f' \sum_i (P'_i U_i)$. To minimize this, we must solve the optimization problem:

$$\begin{aligned}
\underset{P'_i, f'}{\text{minimize}} \quad & f' \sum_i P'_i U_i \\
\text{subject to} \quad & H(\hat{Y}) = \frac{\log_2 |\mathcal{X}|}{f'} \\
& \sum_i P'_i = 1.
\end{aligned} \tag{47}$$

For fixed $f'$, the symbol occurrence probability corresponding to the minimum average cost is

$$P'_i = \frac{1}{N} 2^{-\mu U_i}, \tag{48}$$

where $\mu$ is a constant such that

$$\frac{1}{f'} = \frac{-\sum_i P'_i \log_2 P'_i}{\log_2 |\mathcal{X}|} \tag{49}$$

and $N$ is the normalization factor

$$N = \sum_i 2^{-\mu U_i}. \tag{50}$$

Treating $f'$ as a function of $\mu$, and calculating $df'/d\mu$, we see that $f'(\mu)$ is a monotone increasing function of $\mu$, with $f'(0) = 1$. The optimization problem is then equivalent to:

$$\begin{aligned}
\text{minimize} \quad & F(\mu) = \log_2 |\mathcal{X}| \frac{\sum_i U_i 2^{-\mu U_i}}{\sum_i \mu U_i 2^{-\mu U_i} + N \log_2 N} \\
\text{subject to} \quad & \mu \geqslant 0.
\end{aligned} \tag{51}$$

Calculating $dF/d\mu$, we see that its sign is the negative of the sign of $\log_2 N$. If $U_1 = 0$, then $N = \sum_i 2^{-\mu U_i} > 1$, implying that $F(\mu)$ is a monotone decreasing function on $[0, \infty)$. If $U_1 > 0$, then when $\mu = 0$, we have $N = |\mathcal{Y}|$, so $\log_2 N > 0$. Since $dN/d\mu < 0$ for all $\mu$, we conclude that $F$ will decrease as $\mu$ increases, reaching a minimum at $N = 1$. Beyond that point, $\frac{dF}{d\mu} > 0$.

Thus, the corresponding expansion factor that achieves the minimum total cost is

$$f' = \frac{\log_2 |X|}{-\sum_i P'_i \log_2 P'_i} \tag{52}$$

where $P'_i = 2^{-\mu U_i}$, and $\mu$ is a positive constant satisfying $\sum_i 2^{-\mu U_i} = 1$. The expansion factor of the optimal shaping code is therefore

$$f = f' \frac{H(\mathbf{X})}{\log_2 |X|} = \frac{H(\mathbf{X})}{-\sum_i P'_i \log_2 P'_i}. \tag{53}$$

■

A similar but weaker version of Theorem 8 can be found in [1].

**Remark 6.** If we only apply optimal lossless compression to the source $\mathbf{X}$, the code sequence has a uniform distribution. Therefore, we have $\mu = 0$ and $N = |\mathcal{Y}| > 1$. This implies that simply applying compression to the source data is not the best way to reduce the total wear cost.

### C. Performance of Direct Shaping Codes

Direct shaping codes [8] have parameters $q = 1$, $f = 1$. Suppose the input $\mathbf{X}$ and output $\mathbf{Y}$ both have alphabet size $2^m$, and the codewords have cost vector $\mathcal{U}$. The following theorem characterizes the minimum average wear cost.

**Theorem 9.** *Given the distribution $\tilde{P}$ and cost vector $\mathcal{U}$, the minimum average wear cost achievable with an SLC direct shaping code is $\sum_i \tilde{P}_i U_i$.* □

*Proof:* The proof follows from the strong law of large numbers. Details are given in Appendix C. ■

In general, a direct shaping code will be suboptimal for expansion factor $f = 1$. From Theorem 6, we can determine the exact optimality conditions.

**Corollary 10** *SLC direct shaping codes are optimal with respect to average wear cost if and only if*

$$\tilde{P}_i = 2^{-\mu U_i}, \tag{54}$$

*where $\mu$ is a constant such that $\sum_i \tilde{P}_i = 1$.* □

## IV. CONCLUSION

In this paper, we studied information-theoretic properties and performance limits of a general class of data shaping codes. We determined the asymptotic symbol occurrence probability distribution, and used it to determine the minimum achievable average wear cost for a given expansion factor. We then proved a separation theorem stating that optimal data shaping can be achieved by a concatenation of optimal lossless compression and optimal endurance coding. Using these results, we determined the minimum total wear cost and optimal expansion factor for a data shaping code. As an application, we showed that direct shaping codes are in general suboptimal, and gave the conditions under which optimality can be achieved.

## ACKNOWLEDGMENT

This work was supported by NSF Grant CCF-1619053.

## APPENDIX A
## PROOF OF LEMMA 4

*Proof:* In this proof, we will assume $q = 1$. First, we evaluate the expectation of sequence of random variable $\{N_i(\phi(X_1^{M_l}))\}_{l=1}^{\infty}$. Combining lemma 4 with equation (14), we have

$$\lim_{l \to \infty} \frac{1}{l} E(N_i(\phi(X_1^{M_l}))) = \lim_{l \to \infty} \frac{1}{l} E(N_i(\phi(X))) E(M_l)$$
$$= E(N_i(\phi(X))) \lim_{l \to \infty} \frac{1}{l} E(M_l) \quad (55)$$
$$= E(N_i(\phi(X))) \frac{1}{E(L)}.$$

Similarly, we have

$$\lim_{l \to \infty} \frac{1}{l} E(N_i(\phi(X_1^{M_l - 1})))$$
$$= \lim_{l \to \infty} \frac{1}{l} E(N_i(\phi(X_1^{M_l})) - N_i(\phi(X_{M_l})))$$
$$= E(N_i(\phi(X))) \frac{1}{E(L)} - \lim_{l \to \infty} \frac{1}{l} E(N_i(\phi(X_{M_l}))) \quad (56)$$
$$= E(N_i(\phi(X))) \frac{1}{E(L)},$$

where $\lim_{l \to \infty} \frac{1}{l} E(N_i(\phi(X_{M_l}))) = 0$ follows from lemma 2.

By definition,

$$E(N_i(\phi(X_1^{M_l}))) = \sum_{y_1^l} \sum_{x_1^{M_l} \in \mathcal{G}_\phi(y_1^l)} P(x_1^{M_l}) N_i(\phi(x_1^{M_l})). \quad (57)$$

Since $S_{M_l} > l$, we have $N_i(\phi(x_1^{M_l})) \geqslant N_i(y_1^l)$ and $E(N_i(y_1^l))$ can be bounded as follows

$$E(N_i(Y_1^l)) = \sum_{y_1^l} N_i(y_1^l) Q(y_1^l)$$
$$= \sum_{y_1^l} N_i(y_1^l) \sum_{x_1^{M_l} \in \mathcal{G}_\phi(y_1^l)} P(x_1^{M_l})$$
$$\leqslant \sum_{y_1^l} N_i(\phi(x_1^{M_l})) \sum_{x_1^{M_l} \in \mathcal{G}_\phi(y_1^l)} P(x_1^{M_l}) \quad (58)$$
$$= \sum_{y_1^l} \sum_{x_1^{M_l} \in \mathcal{G}_\phi(y_1^l)} P(x_1^{M_l}) N_i(\phi(x_1^{M_l}))$$
$$= E(N_i(\phi(X_1^{M_l}))).$$

Similarly, $N_i(\phi(x_1^{M_l - 1})) \leqslant N_i(\phi(y_1^l))$ and $E(N_i(Y_1^l))$ is lower bounded by

$$E(N_i(Y_1^l)) = \sum_{y_1^l} N_i(y_1^l) \sum_{x_1^{M_l} \in \mathcal{G}_\phi(y_1^l)} P(x_1^{M_l})$$
$$\geqslant \sum_{y_1^l} \sum_{x_1^{M_l} \in \mathcal{G}_\phi(y_1^l)} P(x_1^{M_l}) N_i(\phi(x_1^{M_l - 1})) \quad (59)$$
$$= E(N_i(\phi(X_1^{M_l - 1}))).$$

Equation (58) and (59) imply that

$$\limsup_{l \to \infty} \frac{1}{l} E(N_i(Y_1^l)) \leqslant \liminf_{l \to \infty} \frac{1}{l} E(N_i(\phi(X_1^{M_l}))) \quad (60)$$
$$= E(N_i(\phi(X))) \frac{1}{E(L)}$$

$$\liminf_{l \to \infty} \frac{1}{l} E(N_i(Y_1^l)) \geqslant \limsup_{l \to \infty} \frac{1}{l} E(N_i(\phi(X_1^{M_l - 1}))) \quad (61)$$
$$= E(N_i(\phi(X))) \frac{1}{E(L)}.$$

Thus we prove lemma

$$P(\hat{Y} = \beta_i) = \lim_{l \to \infty} \frac{1}{l} E(N_i(Y_1^l)) = E(N_i(\phi(X))) \frac{1}{E(L)}. \quad (62)$$

∎

## APPENDIX B
## PROOF OF THEOREM 8

*Proof:* The separation theorem implies that, for a source **X**, we can first apply an optimal lossless compression to get a uniform source $\mathbf{X}'$. Now we consider the optimal $f'$ when endurance coding is applied to $\mathbf{X}'$. The total wear cost is $f \sum_i (P'_i U_i)$. To minimize this, we must solve the optimization problem:

$$\begin{aligned} \underset{P'_i, f'}{\text{minimize}} \quad & f' \sum_i P'_i U_i \\ \text{subject to} \quad & H(\hat{Y}) = \frac{\log_2 |\mathcal{X}|}{f'} \quad (63) \\ & \sum_i P'_i = 1. \end{aligned}$$

For fixed $f'$, the marginal distribution corresponding to minimum average cost is

$$P'_i = \frac{1}{N} 2^{-\mu U_i}, \quad (64)$$

where $\mu$ is a constant such that

$$\frac{1}{f'} = \frac{-\sum_i P'_i \log_2 P'_i}{\log_2 |\mathcal{X}|} \quad (65)$$

and $N$ is the normalization factor

$$N = \sum_i 2^{-\mu U_i}. \quad (66)$$

Treating $f'$ as a function of $\mu$, and calculating $df'/d\mu$, we can see that $f'(\mu)$ is a monotone increasing function of $\mu$, with $f'(0) = 1$. Specifically, replacing the $P'_i$ in (65) by (64) gives

$$f' = \frac{N \log_2 |\mathcal{X}|}{(\sum_i \mu U_i 2^{-\mu U_i}) + N \log_2 N}. \quad (67)$$

The derivative of $f'$ with respect to $\mu$ is

$$\frac{df'}{d\mu} = \frac{\mu \ln 2 \sum_{i<j} 2^{-\mu(U_i+U_j)} (U_i - U_j)^2}{[(\sum_i \mu U_i 2^{-\mu U_i}) + N \log_2 N]^2}, \quad (68)$$

which is easily seen to be positive for $\mu > 0$.

The optimization problem is then equivalent to:

$$\begin{aligned} \text{minimize} \quad & F(\mu) = \log_2 |\mathcal{X}| \frac{\sum_i U_i 2^{-\mu U_i}}{\sum_i \mu U_i 2^{-\mu U_i} + N \log_2 N} \\ \text{subject to} \quad & \mu \geq 0 \end{aligned} \quad (69)$$

Calculating $dF/d\mu$, we see that its sign is the negative of the sign of $\log_2 N$. Specifically, we find that

$$\frac{dF}{d\mu} = -\ln 2 \log_2 |\mathcal{X}| \log_2 N \frac{N \sum_i U_i^2 2^{-\mu U_i} - (\sum_i U_i 2^{-\mu U_i})^2}{(\sum_i \mu U_i 2^{-\mu U_i} + N \log_2 N)^2}. \quad (70)$$

Observe that

$$\begin{aligned} & N \sum_i U_i^2 2^{-\mu U_i} - (\sum_i U_i 2^{-\mu U_i})^2 \\ &= \sum_{i,j} U_i^2 2^{-\mu(U_i+U_j)} - \sum_i U_i^2 2^{-2U_i} - \sum_{i<j} 2 U_i U_j 2^{-\mu(U_i+U_j)} \\ &= \sum_{i<j} (U_i^2 + U_j^2 - 2 U_i U_j) 2^{-\mu(U_i+U_j)} \\ &= \sum_{i<j} (U_i - U_j)^2 2^{-\mu(U_i+U_j)} > 0. \end{aligned} \quad (71)$$

Therefore,

$$\frac{dF}{d\mu} = -\ln 2 \log_2 |\mathcal{X}| \log_2 N \frac{\sum_{i<j} (U_i - U_j)^2 2^{-\mu(U_i+U_j)}}{(\sum_i \mu U_i 2^{-\mu U_i} + N \log_2 N)^2}. \quad (72)$$

The claimed relationship between the sign of $dF/d\mu$ and the sign of $\log_2 N$ is then evident.

It follows that if $U_1 = 0$, then $N = \sum_i 2^{-\mu U_i} > 1$, implying that $F(\mu)$ is a monotone decreasing function on $[0, \infty)$. On the other hand, if $U_1 > 0$, then when $\mu = 0$, we have $N = |\mathcal{Y}|$, so $\log_2 N > 0$. Since $dN/d\mu < 0$ for all $\mu$, we conclude that $F$ will decrease as $\mu$ increases, reaching a minimum at $N = 1$. Beyond that point, $\frac{dF}{d\mu} > 0$.

Thus, the corresponding expansion factor that achieves the minimum total cost is

$$f' = \frac{\log_2 |X|}{-\sum_i P'_i \log_2 P'_i} \quad (73)$$

where $P'_i = 2^{-\mu U_i}$, and $\mu$ is a positive constant satisfying $\sum_i 2^{-\mu U_i} = 1$. The expansion factor of the optimal shaping code is therefore

$$f = f' \frac{H(\mathbf{X})}{\log_2 |X|} = \frac{H(\mathbf{X})}{-\sum_i P'_i \log_2 P'_i}. \quad (74)$$

∎

## APPENDIX C
## PROOF OF THEOREM 9

*Proof:* We make use of the notation and terminology of [5]. For any two symbols $\mathbf{w}_i$ and $\mathbf{w}_j$ in the alphabet $\mathcal{X}$, we assume without loss of generality that $\tilde{P}_i > \tilde{P}_j$. Consider a sequence of i.i.d random variables $X_1^{ij}, X_2^{ij}, \ldots$ with $P(X_1^{ij} = 1) = \tilde{P}_i$, $P(X_1^{ij} = -1) = \tilde{P}_j$ and $P(X_1^{ij} = 0) = 1 - \tilde{P}_i - \tilde{P}_j$. The random variable $X_k^{ij}$ corresponds to the change in distance between symbol $i$ and $j$ at time $k$. The expected value of $X_1^{ij}$ is $\mu^{ij} = \tilde{P}_i - \tilde{P}_j$. Define random variable $S_t^{ij} = \sum_{k=1}^t X_k^{ij}$, and note that $S_t^{ij} > 0$ means $n_i(t) > n_j(t)$. By the strong law of large numbers,

$$S_t^{ij}/t - \mu^{ij} \xrightarrow{a.s.} 0. \quad (75)$$

This means that, for any $\epsilon > 0$, $\{S_t^{ij}/t\}_0^\infty$ is within $\epsilon$ of $\tilde{P}_i - \tilde{P}_j > 0$ for all but finitely many values of $t$, with probability 1. This means that, for any 2 symbols $i > j$, $n_i(t) > n_j(t)$ and, therefore, the dictionary is stable, for all but finitely many values of $t$, with probability 1. Since we are calculating the average cost, this means we only need to consider the contributions at times when the dictionary is stable.

Now, consider a sequence of i.i.d random variables $Y_1, Y_2, \ldots$ with $P(Y_1 = \mathbf{v}_k) = \tilde{P}_k$. The total cost after $t$ steps is the random variable $W_t = \sum_{i=1}^t Y_i$. By the strong law of large numbers,

$$W_t/t - \sum_{k=1}^{2^m} \tilde{P}_k U_k \xrightarrow{a.s.} 0 \quad (76)$$

So the average cost is $\sum_{k=1}^{2^m} \tilde{P}_k U_k$. It can be shown that this conclusion holds even when symbols $\mathbf{w}_i$ and $\mathbf{w}_j$ have equal probabilities, i.e., when $\tilde{P}_i = \tilde{P}_j$.

∎